# A silicon integrated microwave photonic beamformer

Chen Zhu, Liangjun Lu*, Wensheng Shan, Weihan Xu, Gangqiang Zhou, Linjie Zhou, and Jianping Chen

*State Key Laboratory of Advanced Optical Communication Systems and Networks, Shanghai Institute for Advanced Communication and Data Science, Department of Electronic Engineering, Shanghai Jiao Tong University, Shanghai 200240, China*
*Corresponding author: luliangjun@sjtu.edu.cn*

**Optical beamforming networks (OBFNs) based on optical true time delay lines (OTTDLs) are well-known as the promising candidate to solve the bandwidth limitation of traditional electronic phased array antennas (PAAs) due to beam squinting. Here we report the first monolithic 1×8 microwave photonic beamformer based on switchable OTTDLs on the silicon-on-insulator platform. The chip consists of a modulator, an eight-channel OBFN, and 8 photodetectors, which includes hundreds of active and passive components in total. It has a wide operating bandwidth from 8 to 18 GHz, which is almost two orders larger than that of electronic PAAs. The beam can be steered to 31 distinguishable angles in the range of -75.51° to 75.64° based on the beam pattern calculation with the measured RF response. The response time for beam steering is 56 μs. These results represent a significant step towards the realization of integrated microwave photonic beamformers that can satisfy compact size and low power consumption requirements for the future radar and wireless communication systems.** © 2020 Optical Society of America

http://dx.doi.org/xxxx

## 1. INTRODUCTION

In modern radar and wireless communications, microwave phased array antennas (PAAs) play an important role [1, 2]. They use the interference of electromagnetic waves from multiple antenna array elements to generate highly directional beams. Due to the aperture effect caused by the aperture fill time, the traditional electronic phased array radar has a limited bandwidth, which causes beam squinting, that is, the beam directions of different microwave frequencies diverge at an increased steering angle [3]. For example, the instantaneous bandwidth of synthetic aperture radar systems (such as PAZ and Cosmo-SkyMed) in orbit is only several hundred megahertz. With the development of microwave photonics, optical true time delay line (OTTDL) technology has been introduced into phased array radars. OTTDL technology can avoid beam squinting because it has a wide operating bandwidth and can control the entire signal spectrum to steer at the same angle [4, 5]. Meanwhile, OTTDLs have the features of low loss and immunity to electromagnetic interference [6-10]. The outstanding performances of photonics-assisted radars have opened intense research interest ranging from fundamental study to practical applications. A variety of optical beamforming network (OBFN) systems with different kinds of OTTDLs have been proposed, based on highly-dispersive fibers [11-13], fiber Bragg gratings [14, 15], microcombs [16, 17], dense wavelength division multiplexers [18], fast scanning lasers [19], and free-space optics [20-23]. They are mainly focusing on OTTDLs with both large tuning range and fine-tuning resolution. These methods show excellent performance in terms of operating frequency and bandwidth. However, most of these solutions are based on discrete photonic devices, which have the disadvantages of large size, high cost, high power consumption, and low stability.

With the continuous development of integrated photonics, various kinds of integrated OTTDLs have been demonstrated, like cascaded or parallel micro-ring resonators (MRRs) [24-26], multi-path switchable delay lines [27, 28], and grating or photonic crystal delay lines [29], on both silicon and silicon nitride (SiN) platforms. Slow light effect-based OTTDLs can provide continuous delay tuning. However, the delay response is wavelength dependent, and it is very difficult to get uniform delay responses within the operating bandwidth, especially when the signal bandwidth is larger than 10 GHz. On the contrary, switchable delay lines can provide large delays with a broad bandwidth and low temperature sensitivity [30]. Most of the reported OTTDLs are based on the thermo-optic (TO) effect [24-26, 31, 32], because of no extra loss induced upon tuning. However, the response time is in the order of microseconds. To realize fast delay tuning, OTTDLs based on the free-carrier dispersion (FCD) effect have also been demonstrated [28], but it suffers an additional loss due to the free-carrier absorption (FCA). Recently, it has been proposed to build graphene-based OTTDLs to utilize the unique properties of graphene [33, 34]. These OTTDLs show promising performances in both loss and switching time. However, the fabrication processes are more complicated and no experimental demonstration has been reported.

Integrated OBFNs based on MRR binary trees [35, 36] and digital switchable delay lines [37] have also been proposed and demonstrated.

The proof-of-concept demonstration of microwave photonic phased array radars based on the integrated OTTDLs shows the potential in size, weight, and power (SWaP) efficiency [38]. However, only OTTDLs are integrated and all the active functions for electro-to-optic and optic-to-electric conversions are realized with off-chip devices. Besides, due to the additional connecting fibers between different components, extra fiber-based OTTDLs are required to compensate for the delay deviation among multiple channels [35-37]. Efforts have also been devoted to the hybrid assembly of InP and SiN chips to realize a chip-scale microwave photonic beamformer. It requires high-resolution alignment of several photonic chips [39]. Recently, a four-by-one silicon integrated OBFN has been implemented in a coherent photonics-aided receiver for applications in communications satellites [40]. Four Mach-Zehnder delay interferometer (MZDI)-based OTDDLs and a germanium PD are integrated into the same chip. It successively receives two beams modulated with 1 Gb/s quadrature phase-shift keying (QPSK) signal at 28 GHz. However, the total insertion loss of the chip is around 35 dB and the system is still very complex.

In this work, we propose and demonstrate a 1×8 microwave phased array chip based on 5-bit switchable OTTDLs. The chip is fabricated on the silicon-on-insulator (SOI) platform. In addition to the OTTDLs, the chip is also integrated with a high-speed electro-optic modulator and 8 photodetectors (PDs). Our chip shows the advantages of compact size, broad operating bandwidth, and wide angle-scanning range. To the best of our knowledge, this is the first demonstration of a monolithically integrated broadband microwave photonic beamformer.

## 2. CHIP STRUCTURE

The 1×8 microwave phased array chip includes a Mach-Zehnder modulator (MZM), an eight-channel OBFN, and a germanium PD array. Figure 1(a) shows the schematic structure of the chip. The modulator is based on a 1×2 asymmetric single-drive push-pull carrier-depletion MZM. One of the output ports is served as a test port for MZM performance characterization. The length of the modulation arms is 3 mm. The target doping concentrations of the p- and n-doping regions are $\sim 4\times10^{17}$ cm$^{-3}$ and $\sim 1\times10^{18}$ cm$^{-3}$, respectively. The doping profile of the lateral PN junction was simulated with Silvaco TCAD [41], and the change of effective index under various voltages was simulated with COMSOL Multiphysics [42]. The simulation results show that the insertion loss of the modulation arm is 5.36 dB and $V_\pi$ is ~5.88 V. The design of traveling-wave electrodes (TWEs) is crucial for high-speed operation of the modulator. Optimized with COMSOL Multiphysics, the TWE has low microwave attenuation and good phase match between the optical wave and the microwave, leading to a large modulation bandwidth. The cross-section of the MZM arms is illustrated in the inset of Fig. 1(a). The TWE is terminated with an on-chip 50 Ohm resistor. Compared with dual-drive differential MZM, this modulator can effectively reduce the total capacitance because of the serial connection of the PN junctions in two arms, thereby reducing the microwave loss of the traveling wave electrode [43]. Besides, the push-pull driving scheme can effectively avoid the chirp phenomenon and improve the modulation performance. The PD array is composed of eight standard germanium-based vertical p-i-n diodes designed for the 28 Gbps bit rate.

The OBFN is composed of a 2×8 power splitter and eight 5-bit switchable OTTDLs. The power splitter is based on one 2×2 multimode interferometer (MMI) and six 1×2 MMIs. One of the input ports is connected with the MZM, while the other is used as a test port. Each OTTDL includes 6 TO switches based on Mach-Zehnder interferometers (MZIs). The switch element consists of two 2×2 MMI couplers and two waveguide arms with an equal length of 380 μm. One of the arms is integrated with a TiN microheater based phase shifter. A deep air trench is positioned between the two waveguide arms to suppress the thermal crosstalk and improve the thermal tuning efficiency. The air trench has a width of 22 μm and a depth of 120 μm in the silicon substrate without lateral undercut. The distance between two adjacent MZI switches is 300 μm, which is far enough to suppress thermal crosstalk.

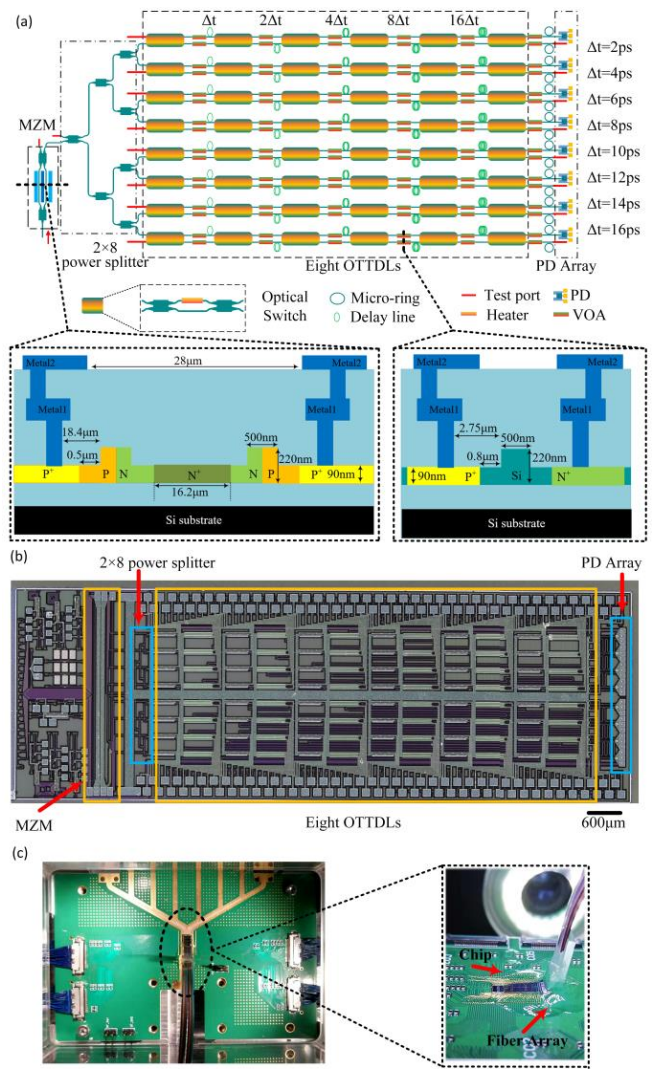

Fig. 1. (a) Schematic of the 1×8 microwave phased array chip. The insets are the cross-sections of the MZM and VOA. (b) Microscope image of the fabricated chip. (c) Picture of the packaged chip.

Two successive optical switches are connected by two waveguides with unequal lengths to form a delay stage, one as a delay waveguide (longer one) and the other as a reference waveguide (shorter one). The time delay of each stage is digitally switched by selecting the shorter or longer optical path. The differential time delay of the $n^{th}$ stage is $2^{n-1}\Delta t$ ($n$=1, 2, ..., 5), where $\Delta t$ is the delay resolution. Therefore, each OTTDL is a 5-bit binary delay line, offering 32 delay states in total. In a one-dimensional PAA, the time delay difference between successive emitters should be increased linearly. By setting the delay difference, the output microwave beam is steered accordingly. In order to get more scanning angles, the delay resolution of each OTTDL is carefully designed. As seen in Fig. 1(a), $\Delta t$ is increased from 2 ps to 16 ps with a step of 2 ps from the first to the last OTTDL. Therefore, as long as the switching states of all the OTTDLs are the same, the OTTDL array can provide a linearly incremental delay varying from 0 ps to 62 ps. In that case, we can have 32 steering angles and the control algorithm is very simple. There are 40 fixed silicon strip waveguide delay lines in various lengths, offering delays from 2 ps to 256 ps. In order to reduce the propagation loss, the waveguide width is designed to be 2 μm for long straight waveguides, while it is tapered down to 500 nm for bending waveguides. The radius of the bending waveguides is 20 μm. The wide and narrow waveguides are connected via 30-μm-long linear tapers. The waveguide group indices changing with widths are simulated with MODE Solution [44] and fitted by a polynomial function. The group delay of uniform waveguides is thus obtained directly. The group delay of the linear tapers is calculated by $\int_L n_g(z)/c\,dz$, where $n_g$ is the group index of the waveguide, $c$ is the light speed in vacuum, and $L$ is the taper length.

Variable optical attenuators (VOAs) based on silicon p-i-n diodes are introduced into all connection waveguides for two main purposes. One is to calibrate the states of the optical switches, and the other is to suppress the crosstalk caused by the limited extinction ratio (ER) of the switches [45]. The cross-section of the VOA is also illustrated in the inset of Fig. 1(a). The highly-doped P+ and N+ regions are separated by 0.8 μm away from the edges of the ridge waveguide to avoid free-carrier absorption loss at zero bias. At the end of each OTTDL, we also insert an MRR and a VOA for each port. MRRs are used to fine-tune the delay to compensate fabrication deviation. The 3-dB passband width of MRRs is about 10 GHz and VOAs are used to adjust the optical power of each delay channel to compensate for the loss difference of OTTDLs. The test ports are reserved for the characterization of key components in the chip.

The chip was fabricated on a 200 mm SOI wafer with a silicon layer thickness of 220 nm and a BOX layer thickness of 2 μm using CMOS-compatible processes in AMF Singapore. The chip is designed for transverse electric (TE) polarization. Figure 1(b) shows the microscope image of the chip. Such a chip contains 170 active elements and hundreds of passive components. The chip footprint is 11.03 mm×3.88 mm. The fabricated chip was wire-bonded to a print circuit board (PCB) so both RF signals and DC control voltages can be applied to the chip. As all the input and output ports are routed to the same side of the chip, a fiber array was vertically coupled with the on-chip grating couplers and fixed by UV-curable adhesive. Figure 1(c) shows the picture of the packaged chip. The coupling loss is ~6.4 dB/facet around 1540 nm wavelength, which can be further improved by using high-efficiency grating couplers or suspended edge couplers [46, 47].

## 3. EXPERIMENTAL RESULTS

### A. Characterization of key elements

Figure 2(a) shows the normalized transmission spectra of the on-chip modulator under various bias voltages. The insertion loss of the modulator is about 12 dB, which is larger than our previous design [48, 49]. We speculate that the large loss comes from the mask misalignment and doping concentration deviation. Due to the unbalanced MZI, the spectra show a periodical response with a free spectral range (FSR) of ~11 nm. The static ER of the modulator gradually decreases from 28 dB to 18 dB when the reverse bias voltage increases from 0 V to 6 V. The modulation efficiency, defined as the product of half-wave voltage and active arm length, is $V_\pi \cdot L$ = 1.35 V·cm. We also measured the electro-optic (EO) S-parameters of the modulator using a vector network analyzer (VNA, Keysight N5247A). As shown by the red line in Fig. 2(b), the 3-dB EO bandwidth is ~6 GHz at 0 V bias. The other lines show the EO responses of a test MZM with the same design parameters using an off-chip 50 Ohm resistor and a 40 GHz RF probe. We can see that the 3-dB EO bandwidth increases to 9.5 GHz at 0 V bias, which indicates that the bandwidth of the on-chip MZM is mainly limited by the RF transmission loss of the PCB and bonding wires. The EO bandwidth increases to 13.8 GHz and 19 GHz at -2 V and -4 V biases, respectively.

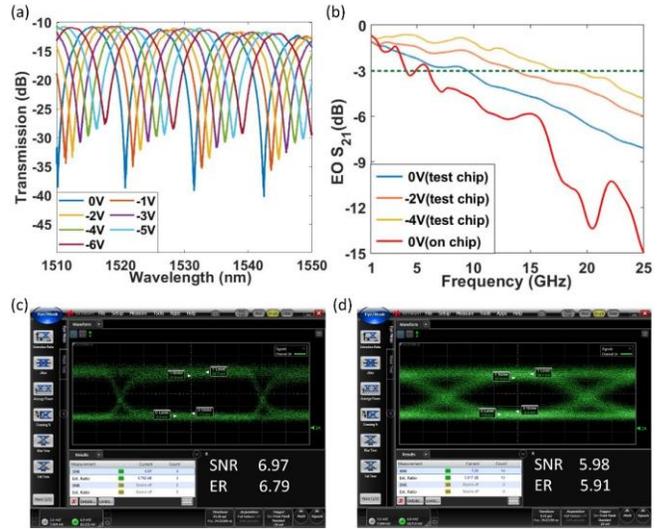

Fig. 2. (a) Transmission spectra of the modulator under various bias voltages. (b) EO $S_{21}$ response of the modulator. (c, d) Measured OOK eye diagrams of the modulator at (c) 5 Gb/s and (d) 20 Gb/s.

We carried out a high-speed on-off keying (OOK) modulation experiment to test the modulation performance of the on-chip modulator. A continuous-wave (CW) light at the 1544.1 nm wavelength was adjusted to TE polarization by a polarization controller (PC) before it was coupled to the modulator. The modulator driving signal is a $2^{23}-1$ pseudo-random binary sequence (PRBS) non-return-to-zero (NRZ) signal, generated by a pulse pattern generator (PPG). In the measurement, the modulator was biased at 0 V. The modulated optical signal was amplified by an erbium-doped fiber amplifier (EDFA) followed by a band-pass filter to suppress the amplified spontaneous emission (ASE) noise. Finally, after optical-to-electrical conversion by a commercial PD (Finisar, XPDV21210RA), the signal was received by a digital sampling oscilloscope (DSO, Keysight, DCA-X 86100D) to test the eye diagrams. Figures 2(c) and (d) show the measured eye diagrams for 5 Gb/s and 20 Gb/s OOK modulations with the on-chip MZM. The ER and the signal-to-noise ratio (SNR) both reach ~6 dB, which indicates that our modulator can support up to 20 Gb/s OOK modulation.

Figure 3(a) shows the measured dark current of the PD as a function of reverse bias voltage. When the reverse bias voltage increases from 2 V to 4 V, the corresponding dark current increases from -14.6 nA to -130 nA, indicating the low noise of the on-chip PD. Figure 3(b) shows the measured photocurrent as a function of optical power under three bias voltages at the 1544.1 nm wavelength. Light was launched from the test port of the first OTTDL channel. Without counting the coupling loss of the grating coupler, the estimated transmission loss of the delay line is ~10.3 dB. The responsivity of the PD is 0.80 A/W at a reverse bias of 2 V, which is used as the bias voltage for delay characterization. All the other PDs have a similar DC response.

We also measured the OOK optical signal detection performance of the on-chip PD. The experimental setup is similar to the above OOK modulation setup. Here, a commercial EO modulator that supports 40 Gb/s modulation was used. The modulated optical signal was amplified to 10 dBm and adjusted to TE mode before it was coupled to the first delay channel of the chip through the test port. The delay line was switched to the shortest delay path. A reverse bias voltage of 2 V was applied to the PD via a bias-Tee. Finally, the converted electrical signal was received by the DSO. Figures 3(c) and (d) show the OOK eye diagrams measured by the on-chip PD. The SNR for 5 Gb/s and 20 Gb/s OOK modulations is above 4.5 dB. It reveals that our PDs are capable of detecting signals from 5 Gb/s to 20 Gb/s.

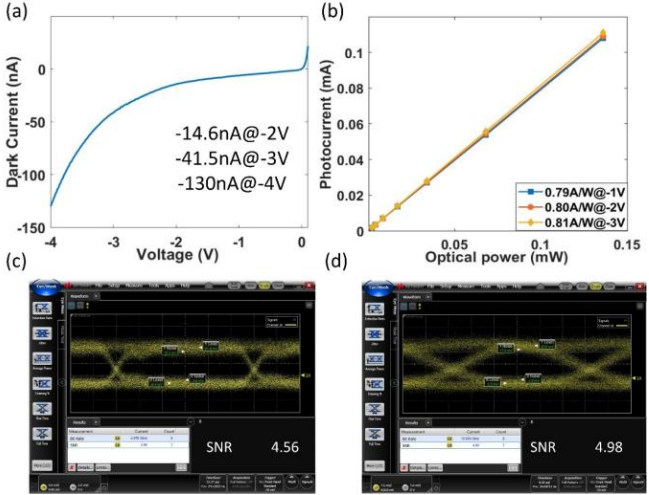

Fig. 3. (a) Dark current of the on-chip photodetector. (b) Photocurrent as a function of input optical power. (c, d) Measured OOK eye diagrams of the photodetector at (c) 5 Gb/s and (d) 20 Gb/s.

Figures 4(a) and (b) show the measured transmission spectra of a single optical switch at the bar and cross state, respectively. The switch shows a broad operating bandwidth from 1515 nm to 1545 nm. At 1544.1 nm wavelength, the insertion loss is 0.38 dB. The crosstalk is lower than -33 dB, which indicates the nearly equal power splitting of the 2×2 MMI. We also measured the switching time by applying a 500 Hz rectangular electrical signal to the TiN microheater. As shown in Fig. 4(c), the peak-to-peak driving voltage is 1.58 V. Figure 4(d) presents the measured optical response. The rise/fall time of the optical signal is ~56 / 16 µs. Therefore, the beam angle can be potentially steered with a maximum response time of 56 µs.

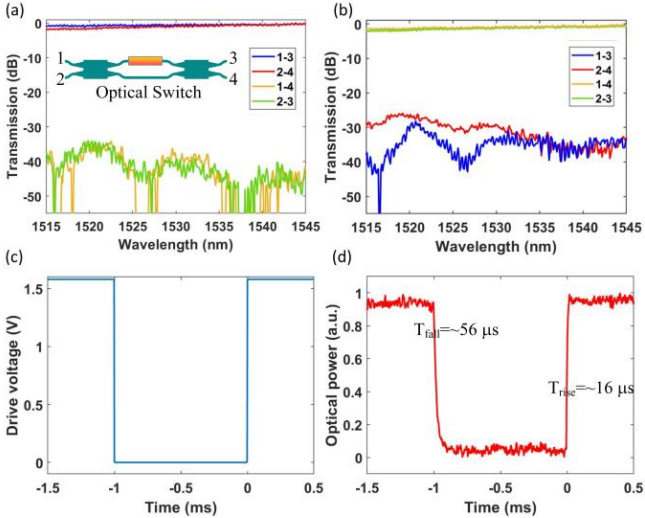

Fig. 4. (a, b) Transmission spectra of a single optical switch at (a) the bar state and (b) the cross state. (c) Temporal waveform of the applied square-wave electrical drive signal. (d) Measured temporal waveform of the optical signal.

### B. Delay performance and beam patterns

For the OBFN, there are total 48 MZI switches. Due to fabrication deviation, the states of the optical switches are random, which induces multi-path interference in the OTTDLs. Typically, extra optical power taps are required to monitor the state of each switch [45]. Here, we calibrated each optical switch with the help of the on-chip VOAs. For a certain delay state, we turn on the VOAs on the unwanted paths so that there is only one transparent optical path without suffering interference. We optimized the applied voltages on the MZI switches by the sequence quadratic program algorithm until the output power monitored by an optical power meter reached the maximum. By selecting different delay states and turning on the corresponding VOAs, the required voltages for all the optical switches were determined and recorded.

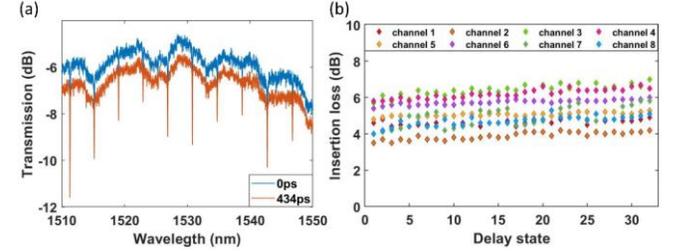

Fig. 5. (a) Transmission spectra for the shortest and longest delay paths in the 7th OTTDL. (b) Measured on-chip insertion loss of 32 delay states for each OTTDL channel.

Figure 5(a) shows the measured transmission spectra of the shortest and longest paths in the 7th OTTDL. The longest path has a delay of 434 ps relative to the shortest one. These two delay paths exhibit similar spectral features but with a loss difference of ~1 dB, indicating the low propagation loss of the waveguides. Due to the low crosstalk of the MZIs, VOAs are unnecessary to suppress light leakage to other unwanted optical paths. We can see that even with the longest delay, there is still no significant ripples in the spectrum. There are several periodical resonance dips with an FSR of ~ 4 nm, coming from the MRR at the end of the OTTDL. The maximum delay of the MRR is 48 ps according to a reference device with the same design. The delay bandwidth is ~0.108 nm (13.5 GHz). According to the beam pattern calculation, an angle steering step of 5° is fine enough for X- and Ku-band applications with the 8-channel beamformer. Therefore, in the beam-forming demonstration, we did not use these MRRs to fine-tune the delay. We chose the non-resonant wavelength at 1544.1 nm as the operating wavelength. The MRRs can further improve the delay accuracy but at the expense of reduced optical bandwidth. Figure 5(b) illustrates the on-chip insertion loss of all the 32 delay states in each OTTDL. The insertion loss varies from 3.5 dB to 7 dB, increasing with the amount of delay. For the shortest delay, the average insertion loss of the eight OTTDLs is ~4.8 dB with a deviation of about ±1 dB. As the insertion loss of an MZI switch is ~ 0.38 dB, the total loss for the six switches in each delay line is 2.28 dB. The additional loss of the OTTDLs mainly comes from the connecting waveguides and the VOAs. The estimated average waveguide propagation loss is thus ~1.3 dB/cm. The loss variation is due to the non-uniformity of optical switches, silicon waveguides, and grating couplers.

Figure 6 depicts the experimental setup for the microwave response measurement of the chip. The input light was from a tunable CW laser with 10 mW optical power. Due to the large insertion loss of the on-chip MZM, we used a commercial modulator instead. The modulated light was coupled to the chip from the input test port of the 2×8 power splitter. The modulator was driven by the RF signal from the VNA. The modulated optical signal was amplified to by an EDFA to compensate for the insertion loss of the modulator and the chip, followed by a bandpass filter to eliminate ASE noise. Then it was adjusted to TE polarization through a PC before coupled to the chip from the test port of the 2×8 power splitter. The injected on-chip optical power is ~12 dBm. A multi-channel programmable voltage source was used to tune the optical switches in the chip to specific states. The optical signal was split into eight channels by the power splitters and delayed by the switchable optical delay lines. Finally, the delayed optical signals were detected by the on-chip PDs and received by the VNA one-by-one for the measurement of microwave signal amplitude and phase.

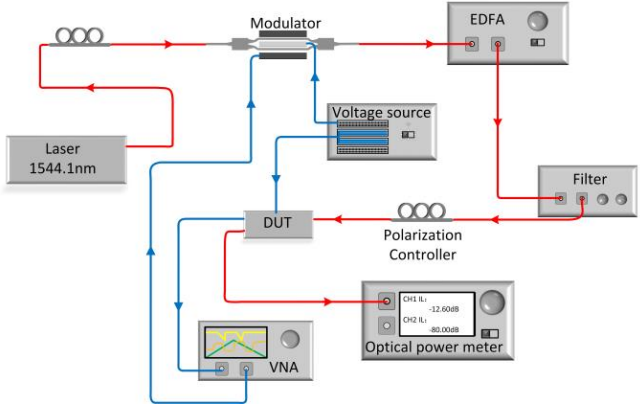

Fig. 6. Experimental setup for microwave transmission response measurement. DUT: device under test.

Figure 7(a) shows the phase response of the 7th channel as a typical result. The phase delay is relative to the shortest one. For all 32 delay states, the phase increases linearly with the microwave frequency. We obtained the group delay response from the frequency derivative of the phase, as illustrated in Fig. 7(b). In the range of 8 GHz ~ 18 GHz (X and Ku bands), the delay response is reasonably flat. The relative delay can be tuned from 0 ps to 434 ps, with a step of 14 ps.

We measured the delay responses of all the 8 OTTDLs for the 32 delay states. As all the delay channels are of similar structures, the phase responses are similar to Fig. 7(a) but with different slopes. We calculated the delay errors between the measured delays and the target ones for all 8 OTTDLs at 16 GHz, as shown in Fig. 7(c). The 8th OTTDL has a relatively larger delay error than the other 7 OTTDLs. The maximum delay error is less than 2.5 ps. The delay deviation may come from the waveguide width deviation due to the fabrication deviation. Figure 7(d) shows the measured microwave transmission ($S_{21}$) response of 32 delay states for all eight OTTDLs at 16 GHz. The transmission varies from -40.2 dB to -47.2 dB, which mainly comes from the loss difference of the OTTDLs. Due to the large insertion loss of the chip, the output RF power from the PD is low. The microwave power feeding to the emitters can be amplified by linear trans-impedance amplifiers (TIAs) or microwave power amplifiers. The possible way to reduce the RF loss is discussed in Section 3.C.

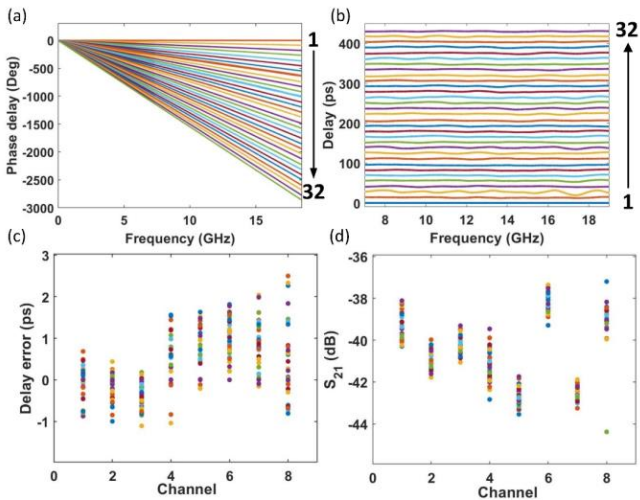

Fig. 7. (a) Microwave phase and (b) group delay responses for all 32 delay states of the 7th OTTDL. (c) Delay errors and (d) Measured $S_{21}$ for all eight OTTDLs at 16 GHz.

With the microwave phase and amplitude responses measured by the VNA, we then calculated the emitting beam patterns formed by the chip. The spacing between adjacent antenna elements was set to 0.94 cm, which is a half wavelength of a 16 GHz microwave signal. The pattern function of the linear phased array is determined by the sum of the individual antenna, which can be expressed as follows [50,51]:

$$F(\theta) = \sum_{i=1}^{8} a_i e^{j(i-1)(\frac{2\pi}{\lambda}d\sin\theta - \phi_i)} \quad (1)$$

where $a_i$ and $\phi_i$ represent the amplitude and phase of the $i^{th}$ antenna element. Substituting the phases and amplitudes of the 8 OTTDLs measured by the VNA into the above formula, we can draw the radiating pattern of the beamforming network in the polar coordinate system.

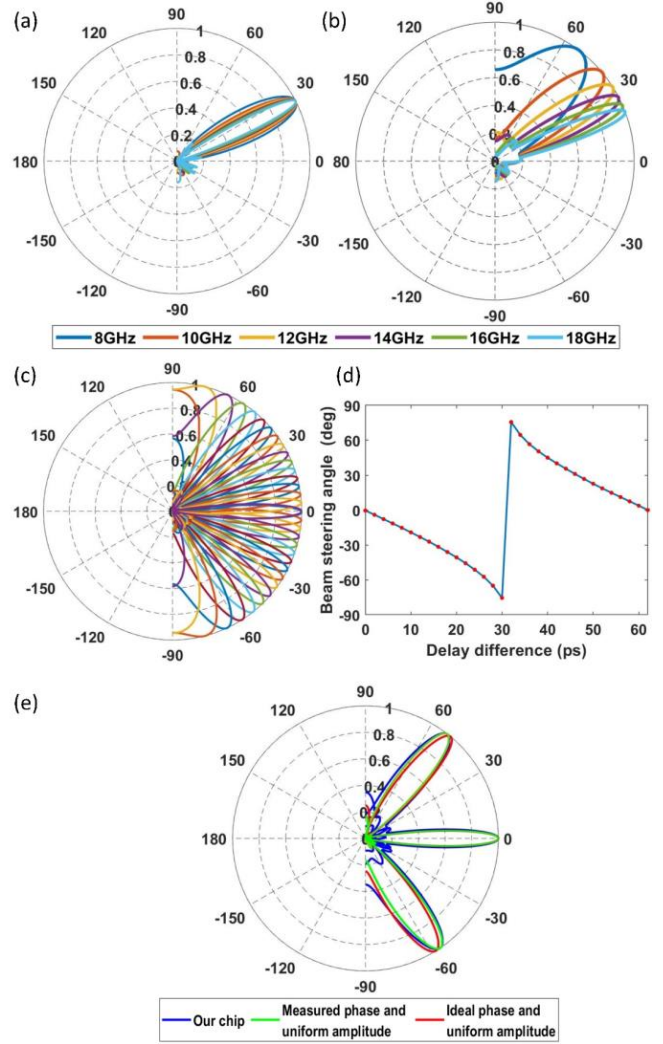

Fig. 8. Calculated beam patterns at 6 RF frequencies for (a) our chip and (b) the electronic beamformer. (c) Calculated beam patterns with 32 tunable steering angles at the 16 GHz frequency. (d) Beam steering angle changes when the delay difference between adjacent emitters increases from 0 ps to 62 ps. (d) Comparison of beam patterns at 16 GHz for three steering angles of 0°, 51.75°, and -55.88°.

Figure 8(a) shows the calculated beam patterns at 6 microwave frequencies from 8 GHz to 18 GHz. The beam was set to point at 26.87°. We also calculated the beam patterns with an electronic PAA, as shown in Fig. 8(b). It is assumed that the phases of all frequencies in the electronic phased array are the same as that of 16 GHz. Other parameters like the spacing between adjacent omnidirectional antenna elements and amplitude are the same as our chip. We can see that our chip can support a large microwave bandwidth (8~18 GHz) without observable beam divergence. In contrast, in a traditional electronic PAA, the beam is squinted from 63.81° to 23.81° when the microwave frequency is increased from 8 GHz to 18 GHz. We can steer the beam angle by setting the optical switches in each column of the OBFN to the same states. Figure 8(c) shows the beam steering patterns of all 32 delay states at 16 GHz. The beam angle was tuned from -75.51° to 75.64° with an average step of 5°. Figure 8(d) presents the beam steering angle changing as a function of incremental delay between adjacent emitters. The incremental delay of the 32nd delay state is 62 ps, corresponding to nearly $2\pi$ phase shift for the 16 GHz microwave signal. Therefore, the beam patterns of the 1st and 32nd delay states are almost overlapped, offering 31 distinguishable steering angles. The 3-dB beamwidth is 11.45 ° at 0°. We also studied the effect of amplitude and phase deviation on the beam shape, as shown in Fig. 8(e). The optical power non-uniformity among eight OTTDLs raises the sidelobe level. We can use VOAs at the end of the OTTDLs to adjust the emitting optical power. The delay error of OTTDLs causes a slight shift of the beam angle. As the delay error is less than 2.5 ps, the maximum angle deviation is less than 2.5°. Therefore, it is still easy to distinguish adjacent beams and not affect the function of our chip. We can also tune the resonances of MRRs at the end of OTTDLs to correct the delay errors if a higher delay accuracy and a larger number of steerable angles are demanded. For example, to compensate the delay deviation of the OTTDLs, a delay tuning of 2.5 ps is required for the MRRs, which brings 27 ps/nm group delay dispersion. Such a distortion could be neglected for a narrow-band microwave signal.

### C. Noise figure and SFDR of the chip

The RF link performance including noise figure and spurious-free dynamic range (SFDR) of the integrated beamformer were also tested and evaluated using the methods in [38, 52]. The experimental setup is similar to that in Fig. 6, except that the input RF signal was generated by an analog signal generator (Keysight N5183B), and the output RF signal was measured by an electrical spectrum analyzer (ESA, Agilent EXA N9010A) with a noise level of -148 dBm. The link gain G is calculated with the ratio of the measured output RF power to the input RF power. The electrical noise power spectral density $P_N$ is retrieved by measuring the RF noise spectrum at a resolution bandwidth of 10 Hz. Therefore, the noise figure is calculated by [38]

$$NF = P_N - G + 174. \qquad (2)$$

Figure 9(a) shows the measured noise figure of the 4th delay channel configured to the shortest and longest delay paths. The noise of the link comes from the relative intensity noise (RIN) of the laser, the ASE noise of the EDFA, and also the thermal noise and shot noise of the PD. Due to the insertion loss of the chip including the loss from the integrated MZM, the fiber-to-chip coupling loss, and the intrinsic power splitting loss, the link gain is below -40 dB, which is similar with [39]. As a consequence, the noise figure is larger than 75 dB. The noise figure increases with RF frequency, caused by the limited bandwidth of both the integrated MZM and the on-chip PD. We also characterized the noise figure of the same delay paths by using a commercial MZM with an insertion loss of 4.8 dB and $V_\pi$ of 2.5 V instead of the silicon MZM. With the superior performance of the commercial MZM, the noise figure is decreased by 12 dB as shown by the blue and pink curves in Fig. 9(a). As the maximum output power from the EDFA is 23 dBm, the generated photocurrent at the on-chip PD is only ~0.3 mA, which essentially limits the noise figure of the link gain. We also measured the noise figure of an RF link bypassing the chip. The E/O and O/E conversions were both performed by the commercial MZM and PD. The received optical power was adjusted by the EDFA to get a similar link gain. The green line shows the retrieved noise figure, valued at 65 dB to 68.5 dB. These results verify that the noise of the on-chip PD is comparable with the commercial PD. We also note that the retrieved $P_N$ is around -148 dBm/Hz, which is limited by the noise level of the ESA. The noise figure can be improved by more than 20 dB if we used a more advanced ESA (e.g., Keysight PXA N9030B with a noise level of -174 dBm). Another main reason for the low RF link gain is due to the 18 dB intrinsic loss from the 2×8 optical power splitter. The noise figure is estimated to be 40 dB for a single-channel delay line without power splitters.

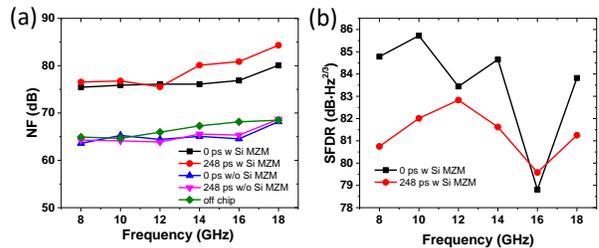

Fig. 9 Measured (a) noise figure and (B) spurious-free dynamic range of the 4th OTTDL for the shortest and longest delay paths in the frequency range of 8 GHz to 18 GHz.

For the SFDR measurement, a standard two-tone test was carried out. The two tones with a frequency interval of 1 MHz were combined by an RF coupler to drive the silicon MZM. After passing the OBFN, the optical signal was converted to the electrical one by the on-chip PD. The $SFDR_3$ was obtained by measuring the fundamental RF tones and the generated third-order intermodulation distortion ($IMD_3$) components on the ESA. Figure 9(b) shows measured $SFDR_3$ of the 4th channel when the time delay is tuned to 0 ps and 248 ps, respectively. The measured average $SFDR_3$ is around 83 dB·$Hz^{2/3}$, covering the 10 GHz frequency range. The difference between these two paths is mainly due to the loss difference. The non-linearity of the link comes from the modulator and the PD. According to our previous work, the $SFDR_3$ of a single-drive push-pull silicon MZM is ~97.7 dB·$Hz^{2/3}$ [49]. The dynamic range can be improved by reducing the insertion loss of the chip.

Both the noise figure and the dynamic range of the chip can be further improved by optimizing the insertion loss of the whole chip. The coupling loss can be reduced to 1.4 dB by using a suspended edge coupler [47]. Recently, we have demonstrated a highly efficient silicon MZM based on a U-shaped PN junction [53]. The $V_\pi$ is reduced to 1.7 V at -2 V bias and the insertion loss of the MZM is lower than 8 dB. Both the low insertion loss and the high modulation efficiency of the MZM increases the microwave link gain of the chip. The insertion loss of OTTDLs can also be reduced by optimizing the optical switches and using ultra-thin silicon waveguides [45]. Besides, the thin waveguides can tolerate high optical power than the regular 220 nm × 500 nm silicon waveguides, as the optical modal field expands more into the oxide cladding. Therefore, the link gain of the beamformer can be improved. Microwave amplifiers like linear TIAs can also be used to increase the link gain of the whole system. The modulator was operated at the quadrature bias point in the measurement. The noise figure can also be improved by optimizing the bias voltage of the modulator [52].

### D. Power consumption of the OBFN

The power consumption of the eight-channel OBFN mainly comes from the thermal phase shifters in the optical switches and VOAs. There are total 48 optical switches in our chip, working either at the cross or the bar state. Table 1 lists the power consumption of the switches. Due to fabrication-induced non-uniformity, it varies across all switches. The power consumption for a $\pi$ phase shift is around 20 mW. The total average power consumption of all optical switches is ~ 1.1 W.

VOAs are introduced into all connection waveguides after the optical switches. In the switching state calibration, VOAs in the unwanted paths are turned on. The applied power is around 50 mW for each VOA. Since the switches have low crosstalk of ~ -33 dB, the influence of multi-path interference between the noise paths and the delay path is negligible. In the measurement, the transmission spectra of the delay lines are almost the same whether the VOAs are turned on or not. Therefore, the VOAs are not used. The VOAs in the last stage are used to adjust the optical power of each channel. As the maximum optical loss difference is 3.5 dB, the maximum power consumption is much less than 350 mW for 7 turned-on VOAs. Therefore, the maximum power consumption of the OBFN is ~1.45 W, including the power from optical switches and VOAs.

Table 1. Electrical power consumption of the optical switches.

| Switch | Elec. power (mW) cross | bar | Switch | Elec. power (mW) cross | bar |
|---|---|---|---|---|---|
| $SE_{11}$ | 3.21 | 24.40 | $SE_{12}$ | 7.05 | 29.26 |
| $SE_{13}$ | 1.63 | 23.32 | $SE_{14}$ | 13.98 | 34.00 |
| $SE_{15}$ | 12.52 | 34.56 | $SE_{16}$ | 14.07 | 24.48 |
| $SE_{21}$ | 10.95 | 32.16 | $SE_{22}$ | 14.38 | 35.58 |
| $SE_{23}$ | 10.63 | 31.84 | $SE_{24}$ | 19.95 | 40.83 |
| $SE_{25}$ | 17.26 | 38.30 | $SE_{26}$ | 12.40 | 34.21 |
| $SE_{31}$ | 20.59 | 41.79 | $SE_{32}$ | 20.99 | 43.12 |
| $SE_{33}$ | 18.55 | 40.87 | $SE_{34}$ | 17.99 | 38.79 |
| $SE_{35}$ | 11.95 | 3.86 | $SE_{36}$ | 7.72 | 28.93 |
| $SE_{41}$ | 5.78 | 28.87 | $SE_{42}$ | 9.75 | 32.27 |
| $SE_{43}$ | 10.82 | 32.76 | $SE_{44}$ | 10.79 | 21.07 |
| $SE_{45}$ | 5.64 | 24.66 | $SE_{46}$ | 13.46 | 34.49 |
| $SE_{51}$ | 18.01 | 39.89 | $SE_{52}$ | 18.04 | 39.31 |
| $SE_{53}$ | 18.26 | 40.22 | $SE_{54}$ | 20.89 | 40.34 |
| $SE_{55}$ | 12.58 | 33.32 | $SE_{56}$ | 10.46 | 29.92 |
| $SE_{61}$ | 6.28 | 28.33 | $SE_{62}$ | 13.11 | 34.89 |
| $SE_{63}$ | 7.90 | 29.26 | $SE_{64}$ | 10.67 | 31.64 |
| $SE_{65}$ | 10.17 | 30.66 | $SE_{66}$ | 19.09 | 40.63 |
| $SE_{71}$ | 16.62 | 38.75 | $SE_{72}$ | 14.89 | 35.97 |
| $SE_{73}$ | 7.00 | 28.95 | $SE_{74}$ | 19.86 | 40.64 |
| $SE_{75}$ | 2.04 | 23.18 | $SE_{76}$ | 16.94 | 37.96 |
| $SE_{81}$ | 18.16 | 39.43 | $SE_{82}$ | 11.54 | 33.52 |
| $SE_{83}$ | 6.11 | 28.67 | $SE_{84}$ | 12.31 | 32.80 |
| $SE_{85}$ | 12.73 | 34.03 | $SE_{86}$ | 10.76 | 31.19 |

We compare our work with the state-of-the-art integrated OTTDLs and OBFN chips as illustrated in Table 2. Although the chip in [17] has the largest number of channels, all the delays are based on optical fibers and controlled by discrete components. Other than that, our chip has a larger number of channels with a more compact chip size and a broader operating bandwidth. Due to the larger delay tuning range of our chip, it has a wider angle-scanning range. Most of all, our chip integrates not only the OTTDLs but also all the E/O and O/E components, which is more compact and robust. As the optical delay is tuned by altering the optical path length by optical switches, the operating bandwidth is much larger, which is mainly limited by the bandwidth of the modulator and PDs. Due to optical power splitting, on-chip optical amplifiers are required to further scale up the beamformer. Another possible way to implement phased array radar systems with more than hundreds of elements is to construct a large beamforming network using the proposed chips as basic units for fine delay tuning. These basic units can be connected by optical fibers, and the optical loss can be compensated by EDFAs.

Table 2. Performance comparison of various integrated microwave photonic beamformers.

| Structure | Integration components level | Platform | Channel number | Footprint (mm$^2$) | Bandwidth (GHz) | Delay range (ps) | Power consumption (mW) | Beam angle range (°) |
|---|---|---|---|---|---|---|---|---|
| CROW [31] | Delay lines | SiON | 1 | 7 | 6.25 | 0~800 | 4000 | - |
| MRR [32] | Delay lines | SOI | 1 | 0.25×0.25 | 10.5 | 0~345 | - | - |
| MZ-TDDL [28] | Delay lines | SOI | 1 | 11.84 | Large | 0~1270 | 1384 | - |
| MRR [54] | Delay lines | SiN | 4 | 36×8 | 8 | 0~139.7 | - | - |
| MRR [37] | Delay lines | SiN | 4 | 32×8 | 8.6 | 0~209 | 2570 | - |
| MRR [56] | Delay lines | SiN | 2 | 2×11 | 5 | 34~252 | - | -28~34 |
| MRR [57] | Delay lines | SOI | 4 | 4.575×0.8 | 2 | 36~200 | 862 | -30~30 |
| Comb [17] | Multi-wavelength source | Thick SiN | 81 | - | 15 | 0~222 | - | -69.7~72.9 |
| AWG [55] | Delay lines | Silica | 8 | 32×71 | 4 | 0~277.83 | - | -52.5~52.5 |
| MZ-TTDL [36] | Delay lines | SiN | 4 | 32×8 | - | 0~22.5 | - | -51~34 |
| MZDI [40] | Delay lines and a PD | SOI | 4 | 6.11×2.88 | - | 0~50 | - | - |
| MZ-TTDL (Our work) | Modulator, PDs and delay lines | SOI | 8 | 11.03 × 3.88 | 10 | 0~496 | 1450 | -75.51~75.64 |

## 4. CONCLUSION

We have demonstrated, to the best of our knowledge, the first monolithic integrated 5-bit 1×8 microwave photonic beamformer chip on the SOI platform, including an eight-channel OBFN, a modulator, and 8 PDs. The output beam is formed and steered by adjusting the switching states of the optical switches. The eight OTTDLs have an insertion loss from 3.5 dB to 7 dB when the relative delay increases from 0 ps to 496 ps. The delay deviation is less than 2.5 ps. It can support X and Ku bands with 8 channels and has a large beam angle tuning range from -75.51° to 75.64°. Such a chip has the advantages of small size (11.03 mm × 3.88 mm), low power consumption (1.45 W), and fast response (56 μs), which is promising for applications in broadband phased array radars for high-resolution imaging and broadband wireless communications. In the future, the chip can be further improved by optimizing the loss of the modulator and integrating on-chip optical amplifiers to compensate for the insertion loss of the chip. The successful implementation of our microwave photonic beamformer marks a significant step forward in the full integration of active and passive components on a single chip and opens up avenues toward real applications of miniature on-chip radar systems.

**Funding.** National Key R&D Program of China (2018YFB2201702, 2019YFB2203203), National Natural Science Foundation of China (NSFC) (61705129, 61535006) and Shanghai Municipal Science and Technology Major Project (2017SHZDZX03).

**Acknowledgment.** We thank AMF Singapore for device fabrication.

**Disclosures.** The authors declare no conflicts of interest